\documentclass[preprint,showpacs,preprintnumbers,amsmath,amssymb,superscriptaddress]{revtex4}

\usepackage{graphicx}

\usepackage{bm}

\newcommand{\pderiv}[2]{\frac{\partial #1}{\partial #2}}

\newcommand{\deriv}[2]{ \frac{d #1}{d#2} }

\begin{document}

\title{Rotational kinetics of absorbing dust grains in neutral gas}

\author{ A.M. Ignatov}\email{aign@fpl.gpi.ru}
\affiliation{General Physics Institute, Moscow, Russia}

  \author{S.A. Trigger}\affiliation{Humboldt University,  Berlin, Germany}
   \author{S.A.Maiorov}\affiliation{General Physics Institute, Moscow, Russia}
  \author{W.Ebeling}\affiliation{Humboldt University,  Berlin, Germany}

\date{\today}

\begin{abstract}
We study the rotational and translational kinetics of massive
particulates (dust grains) absorbing the ambient gas. Equations
for microscopic phase densities are deduced resulting in the
Fokker-Planck equation for the dust component. It is shown that
although there is no stationary distribution, the translational
and rotational temperatures of dust tend to certain values, which
differ from the temperature of the ambient gas. The influence of
the inner structure of grains on rotational kinetics is also
discussed.
\end{abstract}


\maketitle

\section{Introduction}

Recently there was growing interest to the investigation of
composite media, called dusty plasmas, consisting of  aerosol
particles in a gas discharge. Besides numerous industrial
applications dusty plasmas provide ample opportunities to studying
{\sl in situ} phase transitions in the dust component, intergrain
interactions, grain charging etc.

One of the peculiar features of  dusty plasmas is that the average
kinetic energy of the dust component, {\sl i.e.,} its
translational temperature, may be considerably higher than the
temperature of the ambient plasma. In  details the problem was
studied in the  recent papers by Quinn and Goree
\cite{quinn1,quinn2}, where in parallel with the experiments a
model of  the Brownian motion explaining anomalous kinetic
temperature was developed.

The  kinetic description of dusty plasma was discussed in numerous
theoretical studies
 \cite{zagor,schram,tsyt1,tsyt2,tsyt3,sitenko,ignatov}.
  Generally, there are two ways  plasma
particles interact with dust grains: first,
 the scattering of a particle by  grain electric field and,
second, the direct impact of a particle on a grain surface. The
latter process results in grain charging due to the higher
mobility of electrons, it may change grain mass, heat its surface
{\sl etc}. In other words, as  it is well-understood nowadays, the
adequate statistical  description of the dust component should
take into account  inner degrees of freedom, the most important
among which is the grain charge. Kinetic consideration of charging
process shows that absorption of small plasma particles by grains
can result in inequality of the grain temperature and the
temperatures of the light components even for the case of equal
temperatures of electrons and ions\cite{zagor,schram}. To avoid
confusion it should be noted that since the system is open there
is no conflict of  the latter result with thermodynamics.

 Under
the conditions of  experiments on  fine grain synthesis
\cite{childs,gal} or etching \cite{stoffels}  grain mass should
also be treated as a dynamic variable \cite{13,14,15}.
Thermophoretic \cite{jellum} and radiometric \cite{am} forces
provided by inhomogeneous heating of the grain surface may also
play an important role, under microgravity
 especially \cite{morfill}, that may necessitate  inclusion of
  the temperature distribution inside  grains into the kinetic
  description.
  Recent experiments on  rod-like particulates \cite{khrapak}
   demonstrated  complicated rotational motion of dust grains.
   In more general context, the rotation  is also
   of importance for Brownian motion of
 the particles with energy supply \cite{ebeling}.
   Although the rotation of spherical grains is hardly observable
   experimentally, it is of interest since it may influence
    translational motion and
   heating of the grain surface.

   Being not quite complete, the above list of various process
   indicates that the dimension of the phase space required for
   the kinetic description of the dust component may be rather
   large: it tends to a rather large  value of the order of
   twenty.

   The main purpose of  the present paper is to develop the
   kinetic theory of the aerosol component taking into account the mass
   growth and the rotation of grains. Although our main impetus is
   dusty plasma, here we ignore the process of grain charging.
   This may be justified assuming the size of  grains exceeds
   the Debye length of the ambient plasma. In other words, we
   treat the ion component as a neutral gas and neglect the
   influence of electrons. Another reason for accepting this
   simplification
   is that we are able to study the problem both analytically and
   by molecular dynamics simulation: computations with charged moving
   grains are  on the brink of or beyond contemporary  computer capabilities.

   Thus,  we adopt here a following toy
model of the aerosol component absorbing the ambient gas. The dust
component consists of  spherical rotating grains with variable
mass and, consequently, size and moment of inertia. Every atom
hitting the grain surface is absorbed by it, transferring
therefore, its momentum, changing the mass of the grain and its
angular velocity. The process is inelastic since a part of
projectile atom energy is spend for heating the grain surface.  It
is assumed that the size of grains is small compared to the mean
free path of the ambient gas, however the gas distribution
generally depends on the dust component. Our main finding is that
although there is no stationary state of this system, the average
kinetic and rotational energies of dust eventually tend to certain
fixed values, which differ from each other and the temperature of
the ambient gas.

The paper is organized as follows. In Sec.~\ref{scoll} we discuss
the process of the elementary collision.  Microscopic phase
densities and corresponding equations generalizing Newtonian
dynamics are introduced in Sec.~\ref{smpd}, the latter are used in
Sec.~\ref{sfp} to derive the Fokker-Planck equation (\ref{FP})
describing the evolution of the dust distribution. In
Sec.~\ref{stemp} we obtain the homogeneous solution of the
Fokker-Planck equation and evaluate the effective temperatures.
The numeric algorithms are described in Sec.~\ref{salg}, then we
discuss various laws of collision used for simulations
(Sec.~\ref{sncoll}). The results of simulations, which are in
fairly good agreement with analytical theory, are summarized in
Sec.~\ref{sres}.

\section{Analytic theory}

 \subsection{\label{scoll}Elementary collision}

We consider the dust component consisting of a number of spherical
grains of variable masses, $M$, and of mass dependent
 radii, $a(M)$.
Since we are interested in both translational and rotational
degrees of freedom, the state of a grain is described by the
ten-dimensional vector, $\Gamma=(\bm{ R},\bm{  P},\bm{  G}, M)$,
where  $ \bm{ P} $ is the grain's linear momentum and $ \bm{ G} $
is the angular momentum relative its center of inertia, $\bm{ R}$.
 The angles describing the
rotation are irrelevant due to the sphericity of grains. The only
collision process taken into account is the absorption of an
ambient  gas by  grains, i.e., every atom colliding a grain is
assumed to attach to the grain surface transferring  its linear
momentum, angular momentum and mass.

We assume that the process of collision elapses in two stages. At
the first stage the atom attaches to the grain surface. Let $\bm{
r}$ and $\bm{ p}$ be the projectile atom coordinate and,
respectively, the momentum at the instant of collision; evidently,
$|\bm{ r}-\bm{ R}|=a(M)$. The net angular momentum of the
projectile atom and the grain prior to the collision is written as
$\bm{ M}=\bm{ r}\times \bm{ p} + \bm{ R}\times \bm{ P}+\bm{ G}$,
while after the collision $\bm{ M}= \bm{ R}^\prime\times \bm{
P}^\prime+\bm{ G}^\prime$. Since after the attachment the center
of inertia of the grain changes
\begin{equation}
 \bm{ R}^\prime=\frac{ m \bm{
r}+M\bm{ R}}{M+m} \label{ci} \end{equation}
 and $\bm{P}^\prime=\bm{P}+\bm{p}$, the conservation of angular momentum
requires that $$ \bm{ G}^\prime =\bm{ G}+\frac{ (\bm{ r}-\bm{ R})
\times (M \bm{ p}-m\bm{ P})}{M+m}.$$
It should be noted that both $\bm{ G}$ and $\bm{ G}^\prime $ are
independent of the reference frame.

In what follows we combine the above expressions in the convenient
short-hand notation for the process of collision:
\begin{eqnarray}\label{cons}
&&\Gamma \to \Gamma^\prime = \Lambda(\Gamma, \bm{  r}, \bm{ p})
 =(\frac{ m \bm{ r}+M\bm{ R}}{M+m}, \\&& \bm{ P}+\bm{
p}, \bm{ G}+\frac{ (\bm{ r}-\bm{ R}) \times (M \bm{ p}-m\bm{
P})}{M+m}, M+m).\nonumber
\end{eqnarray}

The leap of the center of inertia (\ref{ci}) results in the
non-conservative character of the mapping (\ref{cons}): one can
easily verify that its Jacobian

\begin{equation}
\det \pderiv{\Lambda(\Gamma)}{\Gamma}=\frac{M^3}{(M+m)^3}
\label{jac}
\end{equation}
is less than unity.

 Strictly speaking, the attachment of the atom to the grain surface makes the
 new composite  grain non-spherical.  To
avoid complications with  non-spherical grains we assume that at
the second stage of the collision some inner forces redistribute
the grain mass in such a way that  the grain shapes into a sphere.
Evidently, this does not alter the grain phase variable, $\Gamma$.
Although this assumption seems  a fairly natural simplification,
it may result in  non-physical behaviour because it leaves out the
energy required for the mass redistribution.

 Evidently, the process (\ref{cons}) is
inelastic, that is, a part of the net mechanical  energy is spent
for heating the grain surface, which evidently should be positive.
(The grain heating is of importance if one is going to consider
the aerosol processes like thermophoresis.) Abandoning for a
moment the presumption of sphericity, the amount of heat released
in
 the asymmetric grain  is written as

\begin{eqnarray}\nonumber
\Delta E &=&\frac12  \left\{ \tilde{m}  v^2 + G_i I^{-1}_{ij}G_j
\right.
\\ &-&\left. \left(\bm{ G}+\tilde{m} [ \bm {\rho}\times\bm{ v}]\right)_i
I^{\prime -1}_{ij} \left(\bm{ G}+\tilde{m} [ \bm {\rho}\times\bm{
v}]\right)_j
 \right\}, \label{energy}
\end{eqnarray}
where $\bm{ v} =\bm{ p}/m -\bm{ P}/M$ is the atom velocity in the
grain reference frame, $\tilde{m}=mM/(M+m)$ is the reduced mass,
$\bm{ \rho}=\bm{ r}-\bm{ R}$,  $I_{ij}$ and
$I^\prime_{ij}=I_{ij}+\tilde{m} \left( \rho^2 \delta_{ij}-\rho_i
\rho_j \right) $ are the grain tensors of inertia prior and,
respectively, after the collision. Expanding (\ref{energy}) in
powers of the small mass ratio, $m/M$, we obtain that $\Delta E=
(\bm{ v} -\bm{ \Omega}\times \bm{ \rho})^2/2m\geq 0$, where
$\Omega_i=I^{-1}_{ij}G_j$ is the grain angular velocity. In other
words, the energy conservation does not debar  the mass
absorption.

On the other hand, assuming   the grain  shapes into a sphere, the
tensor of inertia is isotropic and depends on the grain mass only,
$I_{ij}=I(M)\delta_{ij}$, $I^\prime_{ij}=I(M+m)\delta_{ij}$, and
analyzing Eq.~(\ref{energy}) one can  find that $\Delta E$ is
always positive if $dI(M)/dM > a^2(M)$. However, it is easy to
verify for an arbitrary spherically-symmetric mass distribution
that the latter inequality never holds. This means that some
energy is required for the grain  to shape into a sphere and
there exist a certain range of impact parameters, such that  the
available mechanical energy is insufficient  for the mass
redistribution. It should be noted that similar phenomena are also
known in  nuclear physics ({\sl e.g.}, \cite{wu}).

  Simple estimation shows that the assumption of sphericity results
 in the violation of the  energy conservation
for  atoms moving with characteristic velocities of the order of
$a \Omega$.  In the following we assume that the
 energy of rotation is of the order of the gas
temperature and the mass ratio, $m/M$, is small, {\sl i.e. } the
discussed effect is relevant for a very small group of projectile
atoms with velocity less than $\sqrt{m/M} v_T$, where $v_T$ is the
gas thermal velocity.  We ignore the influence of deviation from
 sphericity for this reason. Nonetheless, the discussed effect
may play an important role for small clusters and, perhaps, it
explains the complex structure of fine grains grown in a gas
discharge.

\subsection{ \label{smpd}Microscopic phase densities }
In order to obtain the desired kinetic equation describing the
grain motion we generalize the well-known Klimontovich approach
\cite{klim}.  To describe the whole system the microscopic phase
densities are introduced:

\begin{eqnarray}
N_d(\Gamma,t)  &=& \sum\limits_\alpha
\delta(\Gamma-\Gamma_\alpha(t)), \label{fd}\\ N_n(\bm{ p}, \bm{ r}
,t)  &=& \sum\limits_i \delta(\bm{  p}-\bm{ p}_i(t)) \delta( \bm{
r}-\bm{  r}_i(t)), \label{fn}
\end{eqnarray}
where the indices $\alpha$ and $i$ are used to enumerate grains
and atoms, respectively.

The equations governing the microscopic phase densities are written as
\begin{eqnarray}
\deriv{ N_d(\Gamma,t)  }t  &\equiv&\left( \pderiv{}t + \frac{\bm{
P}}M \pderiv{}{\bm{  R}} +\bm{  F}_d \pderiv{}{\bm{
P}}\right.\nonumber\\ &+& \left. \bm{ K}_d \pderiv{}{\bm{  G}}
\right) N_d(\Gamma,t) , = I_d(\Gamma, \bm{ R})
 \label{eqd}
\end{eqnarray}
\begin{eqnarray}
\deriv{ N_n(\bm{  p},  \bm{  r} ,t)  }t &\equiv& \left( \pderiv{}t
+ \frac{\bm{  p}}m \pderiv{}{\bm{  r}} +\bm{  F}_n \pderiv{}{\bm{
p}} \right) N_n(\bm{  p},  \bm{  r} ,t)\nonumber\\ &=& I_n(\bm{
p}, \bm{ r} ) . \label{eqn}
\end{eqnarray}

Here $F_d$ and $F_n$ are external forces acting upon grains and neutral atoms and $K_d$ is the
external torque. The collision terms in the right-hand sides of
 Eqs.~(\ref{eqd},\ref{eqn}) account for gas
absorption by dust grains; similar terms were introduced in
\cite{schram}. The convenient short-hand forms of these terms are
\begin{widetext}
\begin{eqnarray}
I_d(\Gamma, \bm{  R})= \int d\bm{  p}\;d\bm{  r} \;d\Gamma^\prime
 N_n(\bm{  p},\bm{  r})
N_d(\Gamma^\prime, \bm{  R}) \sigma\left(\bm{  r}-\bm{ R},
\frac{\bm{  p}}{m} -\frac{\bm{  P}^\prime}{M^\prime} , M^\prime
\right)\nonumber\\
\times \left[ \delta\left( \Gamma- \Lambda(\Gamma^\prime, \bm{
r}, \bm{ p}) \right)  - \delta ( \Gamma-\Gamma^\prime) \right] ,
\label{Id}\\
 I_n(\bm{  p},  \bm{  r} )=-\int d\Gamma \;
 \sigma\left( \bm{  r}-\bm{  R}, \frac{\bm{  p}}{m} - \frac{\bm{  P}}{M} , M \right)
  N_n(\bm{  p},\bm{  r})
N_d(\Gamma), \label{In}
\end{eqnarray}
\end{widetext}
where the function $\Lambda(\Gamma, \bm{  r}, \bm{ p})$ is given
by Eq.~(\ref{cons}). The effective cross-section $ \sigma (\bm{
r}, \bm{ v}, M) $ introduced here is

\begin{equation}
\sigma (\bm{  r}, \bm{  v}, M) = -2 (\bm{  r \cdot v })
\theta(-\bm{  r \cdot v }) \delta ( a^2(M)-\bm{ r}^2)
\label{sigma}
\end{equation}
and $\theta(x)$ is the Heaviside's step function.

The collision terms (\ref{Id},\ref{In}) are arranged in a
following way.    Suppose there is a  trajectory described by a
time-dependent radius vector $ \bm{ r}(t)$. If the trajectory
crosses the sphere of radius
 $a$ at the instant $t_0$, i.e., $r(t_0)=a(M)$,
then
\begin{equation}
\sigma(\bm{  r},\dot{\bm{  r}},M)=\delta(t-t_0). \label{prop}
\end{equation}
The step function in (\ref{sigma}) guarantees that  the incoming
intersection point, such that $ \bm{  r} (t_0) \cdot \dot{\bm{
r}}(t_0)<0$,
  is only taken into account.

Substituting the definitions of the microscopic phase densities (\ref{fd},\ref{fn}) to Eqs.
(\ref{Id},\ref{In}) and making use of (\ref{prop})
 one can verify that the collision terms are proportional to
 the sum of $\delta\left(t-\tau_{i\alpha}\right)$,
 where $\tau_{i \alpha}$ is
 the instant of collision of  the $i$-th atom with the $\alpha$-th grain.
At the time intervals between the collisions
Eqs.~(\ref{eqd},\ref{eqn}) describe plain Newtonian dynamics.
However,  at the instant of collision the microscopic phase
densities change abruptly:

\begin{equation}
N_n(\bm{  p},  \bm{  r} , \tau_{i\alpha}+0  )  - N_n(\bm{  p},
\bm{  r} ,\tau_{i\alpha}-0)=
    -\delta(\bm{  p}-\bm{  p}_i)  \delta(\bm{  r}-\bm{  r}_i) , \label{jumpn}
    \end{equation}
that is, the $i$-th atom annihilates.  Simultaneously,  the $\alpha$-th
grain changes its position in the phase space as prescribed by the conservation laws (\ref{cons})

\begin{eqnarray}
N_d(\Gamma,\tau_{i\alpha}+0)  - N_d(\Gamma,
\tau_{i\alpha}-0)=\nonumber\\ \delta(\Gamma-\Lambda(\Gamma_\alpha,
\bm{ r}_i, \bm{ p}_i )) - \delta(\Gamma-\Gamma_\alpha) .
\label{jumpd}
\end{eqnarray}

\subsection{Fokker-Planck equation}\label{sfp}
 The purpose of this section is to expand Eqs.~(\ref{Id},\ref{In}) in powers of the small
 mass ratio, $m/M$.   Averaging over the ensemble and ignoring binary
correlations,
 we may treat $ N_d(\Gamma, t)$ and $ N_n(\bm{  p},  \bm{  r} ,t)$ as
 smooth one-particle distribution functions; in details the procedure was discussed in
  \cite{schram}. Then we integrate over
 $\Gamma^\prime$ in Eq.~(\ref{Id}) and take into account
 Eq.~(\ref{jac}) that results in

\begin{widetext}
\begin{eqnarray}
I_d(\Gamma)&&=\int d\bm{ p} d\bm{ r} \left\{ \sigma\left(\bm{ r},
\frac{M}{M-m} \bm{ v}, M-m \right)
 N_d\left( \bm{ R}-\frac{m}{M}
\bm{ r}, \bm{ P}-\bm{ p}, \bm{ G}-m \bm{ r}\times \bm{ v},
M-m\right) \right. \nonumber\\
&&\left.  \times  N_n\left(\bm{ p},\bm{ R}+\frac{M-m}M \bm{
r}\right)
  -\sigma(\bm{ r},\bm{ v},M) N_d(\Gamma) N_n(\bm{ p}, \bm{
R}+\bm{ r}) \right\},   \label{Id1}
\end{eqnarray}
where
\begin{equation}
\bm{ v}=\frac{\bm{ p}}m -\frac{\bm{ P}}M \label{vel}
\end{equation}
 is the relative velocity. Evaluating Eq.~(\ref{Id1}) we have also
performed the change of variables in the integrand (\ref{Id}):
$\bm{ r}\to \bm{ R}+\frac{M-m}M \bm{ r}$ in the first term and
$\bm{ r}\to \bm{ R}+\bm{ r}$ in the second term.

The next step is the expansion of Eq.~(\ref{Id1}) in powers of the
grain radius, $a(M)$, and mass ratio, $\epsilon =m/M$. Assuming
that $p/P \sim m a v/G \sim \epsilon^{1/2}$ and making use of the
integrals

\begin{eqnarray}
\int d \bm{ r} \sigma(\bm{  r}, \bm{  v}, M) &=&\pi a^2(M) v,
\\
\int d \bm{ r} \sigma(\bm{  r}, \bm{  v}, M) \bm{  r} &=&-\frac23
\pi a^3(M) \bm{  v}, \\
 \int d \bm{ r} \sigma(\bm{ r}, \bm{  v},
M) r_i r_j &=&\frac\pi4   a^4(M) v \left( \delta_{ij}+\frac{v_i
v_j}{v^2} \right)    ,
\end{eqnarray}
we finally arrive at the Fokker-Planck equation for $N_d(\Gamma)$

\begin{eqnarray}
\deriv{N_d( \Gamma) }t = \pderiv{}{P_i} \left\{ - s_i N_d( \Gamma)
+\kappa_{ij}\pderiv{N_d(\Gamma)}{P_j}
 \right\}+\pderiv{}{G_i}
\left(\eta_{ij} \pderiv{ N_d( \Gamma)}{G_j}\right)\nonumber \\
+\pderiv{}{R_i} \left\{ \gamma_i N_d(\Gamma) - \varepsilon_{ijk}
\sigma_k \pderiv{N_d(\Gamma)}{G_j} -\Pi_{ij}
\pderiv{N_d(\Gamma)}{P_j}
 \right\}
-\pderiv{J N_d( \Gamma)}M ,  \label{FP}
\end{eqnarray}
\end{widetext}
where $ \varepsilon_{ijk} $ is the unit antisymmetric tensor.
Here the following kinetic coefficients are introduced:
\begin{eqnarray}
J(\Gamma)&=&\pi a^2(M) m\int d \bm{  p}\; v N_n(\bm{  p},\bm{ R}),
\nonumber
\\
s_i(\Gamma)&=&\pi a^2(M)\int d \bm{  p}\; v  p_i  N_n(\bm{ p},\bm{
R}), \nonumber   \\
 \kappa_{ij}(\Gamma)&=& \frac12 \pi a^2(M)  \int d
\bm{  p}\; v p_i p_j N_n(\bm{ p},\bm{ R}),\nonumber \\
\gamma_i(\Gamma) &=& \frac{2m}{3M} \pi a^3(M)  \int d \bm{  p}\;
v_i N_n(\bm{ p},\bm{ R}), \label{coefs}
\\
 \eta_{ij}(\Gamma) &=& \frac{m^2}8 \pi a^4(M)  \int d \bm{  p}\; v   \left(
\delta_{ij} v^2-v_i v_j \right)
   N_n(\bm{  p},\bm{  R}), \nonumber \\
\sigma_i(\Gamma) &=&\frac{m^2}{4M} \pi a^4(M) \int d \bm{  p}\; v
v_i N_n(\bm{  p},\bm{  R}), \nonumber \\
\Pi_{ij}(\Gamma) &=&  \frac{2m}{3M} \pi a^3(M) \int d \bm{  p}\;
v_i p_j N_n(\bm{ p},\bm{  R}) \nonumber
\end{eqnarray}
and $\bm{ v}$ is given by Eq.~(\ref{vel}).

 The physical meaning of
the most of the coefficients (\ref{coefs}) and corresponding terms
in Eq.~(\ref{FP}) is fairly obvious. $J$ is the mass flow at the
grain surface, the last term in Eq.~(\ref{FP}) provides the mass
growth of the dust component. The coefficient $s_i$ is the drag
force acting upon a grain,   the quantities $\kappa_{ij}$ and
$\eta_{ij}$ characterizes the diffusion in the momentum space. The
term proportional to $\Pi_{ij}$ is just the Archimedean force
 in a non-uniform gas.

Since in the process of collision the angular momentum transferred
to the grain is independent of its angular velocity, there is no
drag torque analogous to the first term in Eq.~(\ref{FP}). This is
the evident consequence of the adopted model. The drag torque may
appear if one takes into account the non-sphericity of grains or
inelastic scattering of atoms by the grain surface. Within the
present model the drag torque is provided by spatial gradients and
it is characterized by   the coefficient $\sigma_i$.  Since in the
low velocity limit, $P/M\ll p/m$, $\sigma_i$ is proportional to
$s_i$, the drag torque may appear if the curl of the drag force is
non-zero.

Of  interest is the term proportional to $\gamma_i$ in
Eq.~(\ref{FP}). It is non-vanishing  if the grain moves relative
to the ambient gas and arises due to the change of the center of
inertia in the process of collision (\ref{ci}) or, in other words,
$\gamma_i$ describes the migration of the center of inertia due to
the asymmetric bombardment of the grain by gas atoms.  Comparing
the left-hand side of (\ref{eqd}) with this term we see that
$\gamma_i$ appears as an addition to the grain velocity, $\bm{
P}/M$.   It often happens that the ambient medium impose some
forces at the particulate, an example is the drag force, $s_i$. In
this case one may say that the medium modifies the second Newton's
law. Here we face with the example of the first Newton's law
altered by the ambient gas.

By the order of magnitude the drag force, $s_i$, is proportional
to $\epsilon^{1/2}$, the effective torque, $\sigma_i \sim
\epsilon^{3/2}$,  all other coefficients are of the order of
$\epsilon$.

 In the lowest-order approximation  the
  kinetics of the gas component is  reduced to the
  absorption. The corresponding kinetic equation is readily
  obtained from Eq.~(\ref{In}):

 \begin{equation}
 \deriv{ N_n(\bm{  p},  \bm{  r} ,t)  }t
 =-\int d\Gamma\; \pi a^2(M) \frac{p}m N_n(\bm{  p}) N_d(\Gamma)+
 I_a (\bm{  p},  \bm{  r}),
\label{abs}
\end{equation}
where $ I_a (\bm{  p},  \bm{  r})$ stands for other dissipative
processes, which were excluded from the above derivation. These
may be, {\sl e.g.,} interatomic collisions, gas creation by
external source {\sl etc.}

\subsection{\label{stemp}Effective temperature}

Suppose that the gas distribution,  $N_n(\bm{  p}) $, is
homogeneous and isotropic. Since we assume that  the grain
velocity is small compared to the gas thermal velocity,  we
neglect the momentum dependence in all kinetic coefficients
(\ref{coefs}) but the drag force, $s_i$, for which the first order
term of expansion in powers of $\bm{ P}/M$ should be kept.
Therefore, there are only three non-zero kinetic coefficients
(\ref{coefs}), which are expressed in terms of mass flow, $J$, and
normalized energy, $\alpha$:

\begin{eqnarray}
s_i&=&-\frac{J}{3M}P_i \nonumber \\
 \kappa_{ij}&=& \delta_{ij} J
\alpha \label{coefs1} \\ \eta_{ij}&=& \delta_{ij} \frac12 a^2(M) J
\alpha, \nonumber
\end{eqnarray}
where
\begin{equation}
\alpha= \frac{a^2(M)}{6m J}  \int d \bm{  p}\; p^3 N_n(\bm{
p}).\label{eps}
\end{equation}

The Fokker-Planck equation (\ref{FP}) is now reduced to

\begin{eqnarray}
\pderiv{N_d(\Gamma)}t &=&\pderiv{}{P_i}\left( \frac{J}{3M}P_i
N_d(\Gamma)+\alpha J \pderiv{N_d(\Gamma)}{P_i} \right)\nonumber
\\
&+&\frac12 \alpha a^2(M)J \frac{\partial^2 N_d(\Gamma)}{\partial
G_i \partial G_i} -\pderiv{J N_d(\Gamma)}M. \label{FPI}
\end{eqnarray}

 Suppose there is no dispersion over the grain mass.
  Then we seek the solution to Eq.~(\ref{FPI}) in the form of
\begin{equation}
N_d(\Gamma)=\delta(M-\mu(t)) f(\bm{ P},\bm{ G},\mu(t)).
\label{anz}
\end{equation}

Evidently, if all the grains are of the same mass and of the same
mass growth rate, we are able to use the current value of mass,
$\mu(t)$ for parameterizing the temporal evolution in the phase
space.  Substituting the latter  ansatz to Eq.~(\ref{FPI}) we find
that
\begin{eqnarray}
\deriv{\mu(t)}t&=&J \label{mu}\\
 \pderiv{f}\mu&=&  \frac{1}{3\mu}\pderiv{P_i
 f}{P_i}+\alpha \Delta_{\bm{ P}} f+
 \frac12 a^2(\mu) \alpha \Delta_{\bm{ G}} f,
 \label{FPR}
 \end{eqnarray}
 where $ \Delta_{\bm{ P}}$ and $ \Delta_{\bm{ G}}$ stand for the
 Laplacian operators acting on the corresponding variables.

It should be noted here that  the normalized energy,  $\alpha$, is
formally independent of mass,  $M$. However, it may depend on
$\mu$ due to the possible time variation of the atom distribution,
$N_n(\bm{ p},t)$.

 Eq.~(\ref{FPR}) is reduced to the diffusion equation by
 changing the variable $\bm{ P}\to \bm{ x}=\mu^{1/3}\bm{ P}$. The
 latter is readily solved resulting in
 \begin{eqnarray}
f(\bm{ P},\bm{ G},\mu)=\int d\bm{ P}^\prime d\bm{ G}^\prime
\frac{\mu f_0(\bm{ P}^\prime, \bm{ G}^\prime)}{(2\pi)^3
(k_1(\mu)k_2(\mu))^{3/2}} \nonumber\\
\times \exp\left[-\frac{( \mu^{1/3}\bm{ P}-\mu_0^{1/3}\bm{
P}^\prime)^2}{2 k_1(\mu)} -\frac{(\bm{ G}-\bm{ G}^\prime)^2}{2
k_2(\mu)}\right],\label{sol}
\end{eqnarray}
where
\begin{eqnarray}
k_1(\mu)&=&2\int\limits_{\mu_0}^\mu \alpha(\mu)\mu^{2/3}d\mu,
\nonumber\\
k_2(\mu)&=& \int\limits_{\mu_0}^\mu \alpha(\mu) a^2(\mu)d\mu,
\nonumber
\end{eqnarray}
$\mu_0$ is the value of mass at $t=0$ and $f_0(\bm{ P},\bm{ G})$
is the corresponding initial distribution.

As it follows from Eq.~(\ref{sol}),  $f(\bm{ P},\bm{ G},\mu)$
eventually tends to the Maxwellian distribution. In order to
obtain the parameters of asymptotic distribution one have to
evaluate the momenta of  $f(\bm{ P},\bm{ G},\mu)$. First, it is
readily checked that the dust density, $n_d=\int d \bm{ P} d\bm{
G} f(\bm{ P},\bm{ G},\mu)$, is independent of $\mu$. Then, we
evaluate the  linear momentum dispersion, $\Delta_t(\mu)$, and the
angular momentum dispersion, $\Delta_r(\mu)$:

\begin{eqnarray}
\Delta_t(\mu)&=&\frac1{3n_d} \int d \bm{ P} d\bm{ G} P^2 f(\bm{
P},\bm{ G},\mu)\nonumber \\ &=&
\left( \frac{\mu_0}{\mu}\right)^{2/3} \Delta_t(\mu_0)+
\frac{ k_1(\mu)}{\mu^{2/3}} ,\label{deltat}\\
\Delta_r(\mu)&=&\frac1{3n_d} \int d \bm{ P} d\bm{ G} G^2 f(\bm{
P},\bm{ G},\mu)\nonumber\\ &=& \Delta_r(\mu_0)+
k_2(\mu).\label{deltar}
\end{eqnarray}

Explicitly, the asymptotic distribution is given by
\begin{eqnarray}
f_{\infty}(P,G,\mu)= \frac{n_d}{(2 \pi)^3 (\Delta_t (\mu)
\Delta_r(\mu))^{3/2}} \nonumber \\ \times \exp \left(-\frac{P^2}{2
\Delta_t(\mu)}-\frac{G^2}{2 \Delta_r(\mu)}\right). \label{anz1}
\end{eqnarray}
It is a matter of direct substitution to verify that the latter
distribution  satisfy Eq.~(\ref{FPI}) subject to relations
(\ref{deltat},\ref{deltar}).

 Now we introduce the
translational temperature, $T_t(\mu)=\Delta_t(\mu)/ \mu$, and the
rotational temperature, $T_r(\mu)=\Delta_r(\mu)/I(\mu)$, where
$I(\mu)$ is the moment of inertia of a grain.  Hereafter   we will
ignore the dependence of the normalized energy, $\alpha$, on
$\mu$: this will be the case of particular examples discussed
below. Then, as it follows from (\ref{deltat}) the translational
temperature tends to a fixed value
\begin{equation}
T_t |_{\mu \to\infty} \to  \frac65  \alpha \label{Tt}
\end{equation}

The asymptotic value of the rotational temperature is determined
by the dependence of its moment of inertia on the mass, i.e., by
the inner structure of a grain. Of interest are two cases. Suppose
that the grain radius is independent of its mass: this will be
referred to as a spongy grain. Then, $I=\frac25 \mu a^2$ and
Eq.~(\ref{deltar}) results in
\begin{equation}
T_r |_{\mu \to\infty} \to  \frac52 \alpha. \label{Tr1}
\end{equation}

Another case is a solid or, more general, a fractal grain with
$\mu \propto a^D$, where $D$ is its fractal dimension. Since the
mass density is proportional to $r^{D-3}$, the moment of inertia
is
\begin{equation}
I=\frac23 \frac{D}{D+2} \mu a^2(\mu). \label{mi}
\end{equation}

 It should be noted that strictly speaking our derivation of kinetic equations
 is applicable to  solid ($D=3$) grains only.
However, the rotational temperature for fractal grains turns out
to be independent of the fractal dimension:

\begin{equation}\label{Tr2}
 T_r |_{\mu \to\infty} \to  \frac32 \alpha.
\end{equation}

Now we turn to evaluation of the normalized energy,
 $\alpha$ (\ref{eps}).
  Although generally the solution of the kinetic equation (\ref{abs})
is time-dependent and can deviate from the Maxwellian
distribution,  there are reasons to ignore this deviation. Suppose
as an example, that the collision term $I_a$ in Eq.~(\ref{abs}) is
represented by a sum of the Boltzmann integral describing
interatomic collisions and let's assume some source term balancing
the gas loss. Then, with the dominating role of elastic
interatomic collisions, the deviation from Max\-wel\-lian
distribution becomes negligible.  Assuming  that $N_n$ is the
Maxwellian distribution with the temperature $T_0$, we get
$\varepsilon = 2/3 T_0$, i.e.

\begin{eqnarray}
T_t &\to& \frac45 T_0 \label{trm} \\
 \label{rotm} T_r &\to&
\left\{
\begin{array}{cc}
  \frac53 T_0 ; &  \text{spongy grain} \\
  T_0; &  \text{fractal grain}
\end{array}
\right.
\end{eqnarray}

Of interest is that the kinetic dust temperature obtained under
the same approximations but in neglecting the mass growth
\cite{schram} is $2\, T_0$. Therefore, the mass growth results in
appreciable cooling of the dust component.

At lower gas pressure the interatomic collisions are negligible
and we have to take into account the deviation of the ambient gas
distribution from Maxwellian. Suppose there is some bulk source of
the Maxwellian gas, i.e. $I_a$ in Eq.~(\ref{abs}) is given by
\begin{equation}
I_a(p) = \nu_0 f_M(p)= \nu_0 \frac{n_0}{ (2 \pi m T_0)^{3/2}} \exp
\left( -\frac{p^2}{2 mT_0} \right). \label{source}
\end{equation}

Assuming that the rate of the dust mass growth (\ref{mu}) is
smaller than the rate of gas creation, i.e., $ m n_0/\mu n_d \ll
1$, we neglect the time derivative in (\ref{abs}), that results in

\begin{equation}\label{da}
N_n(p)= \frac{m \nu_0}{ \xi p} f_M(p), \end{equation}
 where $\xi=
\int d\Gamma \pi a^2(M) N_d(\Gamma)$. Obviously, this distribution
differs from Maxwellian due to the accumulation of slow atoms.
Evaluating the integrals we get $\varepsilon=\frac12 T_0$, that
is, in the case of high dust density

\begin{eqnarray}
T_t &\to& \frac35 T_0, \label{tra} \\
\label{rota}
 T_r &\to& \left\{
\begin{array}{cc}
  \frac54 T_0 ;  & \text{spongy grain} \\
  \frac34 T_0; & \text{fractal grain}.
\end{array}
\right.
\end{eqnarray}

The main results of this section may be summarized as follows. The
momentum distribution of the dust component absorbing the ambient
gas tends to Maxwellian. Although the average value of momenta are
always growing, the corresponding temperatures tend to certain
fixed values, which differ from each other. Moreover, the
rotational temperature depends on the inner structure of dust
grains.

 In ignoring the mass growth, {\sl i.e.,} the last term in
Eq.~(\ref{FPI}), the dust translational temperature in the
Maxwellian gas tends to $2 T_0$ \cite{schram}. However,  due to
the absence of the friction torque within the adopted model,
there is no steady behaviour with respect to the rotational
degrees of freedom.

 These conclusions were gathered assuming the
mass dispersion is negligible. However, the same inferences also
follow from the general non-stationary solution of
Eq.~(\ref{FPI}), which may be obtained in a similar way but is too
cumbersome to be adduced here.

\section{Numeric simulations}
\subsection{\label{salg}An algorithm}

The simulations of the Brownian kinetics of a single grain were
performed in a following way. The computational area was a
three-dimensional cube of unit length on edge in contact with
unbounded equilibrium gas. This contact was simulated by point
atoms of unit mass, which were randomly injected inside the cube
from all of its faces and could freely leave the computational
area. For each atom leaving the cube, another atom with the random
velocity was injected from the random point of the random cube
face. The distribution function of the injected atoms was
semi-Maxwellian.

 It was verified that in the absence of dust grains the bulk
distribution inside the cube was Maxwellian with the prescribed
temperature, $T_0$. The average number of atoms depended on the
thermal velocity; in most runs it fluctuated around 10000. Since
there were no forces acting upon atoms, their trajectories were
straight lines.

 The grain was represented by a movable sphere of a radius small compared to
  the cube edge, typically, the initial size was $a_0=0.01$.
   If an atom hit the grain surface then it transferred a part of its momenta to
   the grain according to some prescribed rules, which are
   discussed below. In the case of absorbing grain,
   no new atom was injected into the cube after the collision,
    that resulted in some reduction of density.
   Since the grain was initially small,
   a very little part of atoms could actually experience the collision.
   The equations of motion for both translational
   and rotational degrees of freedom of the grain were solved; the
   time step was small compared to the average time between the
   collisions.

    It was observed that some
spurious force arose when the grain approached  the faces of the
cube. To minimize the influence of this computational effect the
grain was confined near the center of the cube with the help of
the auxiliary spherically-symmetric parabolic potential well. The
parameters of the well were chosen in such a way that even with
the kinetic energy of about 100 $T_0$ the grain could not approach
the faces of the cube. Evidently, such a confinement should result
in multiplication of the grain distribution by a Boltzmannian
factor and could not alter the distribution over the kinetic
energy. Moreover, in application to the dusty plasma the
confinement appears in a natural way. The imposed auxiliary
potential did not influence the motion of atoms.

It is worth noting that although the mass of the grain could be as
large as several millions of atom masses, its mobility was crucial
for our simulations. It was the temptation to avoid the solution
of the grain equations of motion by simply counting down the
energy and momentum transferred to the immobile grain. However,
this way led nowhere: the grain temperature was permanently
increasing without any saturation.

The main goal of our simulations was to accumulate enough data in
order to reconstruct the grain distribution function over its
kinetic and rotational energy. The translational or rotational
energy axis, say, $0<E<20\,T_0$, was split in a number of
sub-bands (usually, 50). Two methods of averaging were used.
First, we could trace the energy variation of a single grain and
evaluate the time it spent in each energy sub-band. Then, these
time intervals were summed up resulting in time-averaged
distribution function.

Another method of averaging was the simulation of the canonical
Gibbs ensemble. Initially, the grain was situated at the center of
the cube, its rotational and kinetic energies were chosen randomly
using the random-number generators of Maxwellian distributions
with corresponding initial kinetic, $T_{t0}$, and rotational,
$T_{r0}$, temperatures. The  evolution of the grain energies was
recorded for the time period $0<t<t_{max}$. The obtained
dependence represented a single sample from the Gibbs ensemble.
The whole procedure was repeated many times with varied random
initial energies but the same initial temperatures.   By counting
down a number of samples in each energy sub-band for a given time
instant we were able to reconstruct the time evolution of the
energy distribution function.

 It should be noted that reconstruction of the
distribution over both kinetic and rotational energies,
$f(E_{t},E_r)$ requires too many samples or too long integration
time. For this reason we could evaluate only the distribution over
kinetic, $f(E_{t})$, or rotational, $f(E_r)$,  energy separately.

\subsection{Atom-grain collisions} \label{sncoll}

As it was mentioned, for test (and fun) purposes we have used
various laws of interaction of an atom with the grain surface.

{\em Specular  reflections.} The atom experienced the specular
reflection in the grain reference frame:

\begin{eqnarray}
\bm{ p}^\prime&=& \bm{ p}-\Delta\bm{ p}, \nonumber\\
 \bm{ P}^\prime&=& \bm{ P}+\Delta\bm{ p}, \label{mirror}\\
 \bm{ G}^\prime&=&\bm{ G},\nonumber \\
 M^\prime&=&M, \nonumber
 \end{eqnarray}
where the transferred momentum is
 $$ \Delta\bm{ p}=2
\bm{ n} \left( \bm{ n}\cdot \bm{ p}-\frac{m}{M}  \bm{ n}\cdot \bm{
P}\right)$$ and $\bm{ n}=(\bm{ r}-\bm{ R})/a$ is a unit vector.

{\em Absorbing grain.} The grain momenta were changed according to
Eq.~(\ref{cons}). We were able to use various dependencies $a(M)$.

The {\em ``cold grain''} is also described by Eqs.~(\ref{cons})
but its mass remained unchanged;  the same law of collision was
used in \cite{schram}. Physically, this corresponds to the diffuse
scattering with the complete energy accomodation at the cold
surface, when the net atom momentum is transferred to the grain
and the influence of the scattered atom is negligible.

\subsection{\label{sres}Simulation results}

\begin{figure}
\includegraphics{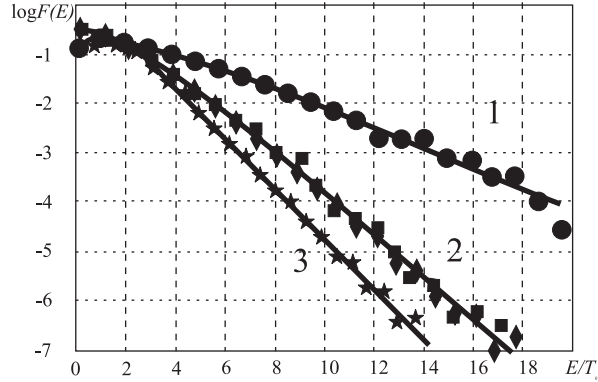}
\caption{\label{fig1} Energy distributions for various kinds of
atom-grain interactions. The distributions over kinetic energy are
plotted for the cold grain ($ \bullet$), for the specular
reflections given by Eqs.~(\ref{mirror}) ($\blacksquare$) and for
absorbing grain ($ \bigstar$). The distribution over rotational
energy is shown for the case of absorbing grain ($\blacklozenge$).
 The solid curves correspond to Maxwellian distributions with
 $T/T_0=2$ (1), $T/T_0=1 $ (2) and $T/T_0=4/5$ (3). }
\end{figure}
The simulations with specular reflections (\ref{mirror}) were used
to check the overall operation of the code. Both averaging methods
described above were used. It was observed that for the relatively
heavy grain  ($M/m=100$) after some thousands of collisions the
distribution function over the kinetic energy eventually tended to
the Maxwellian distribution.
 The dust temperature was equal to $T_0$ with the accuracy less
 than 1\%, as expected for the system in the thermodynamic
 equilibrium. There was no relaxation in rotational degrees of
 freedom because there was no coupling between translational and
 rotational motions.

 The simulations with absorbing grain were performed with the
  solid grain, {\sl i.e.} $a(M)\propto M^{1/3}$.  Although the
  computer facilities allowed
  us to monitor the motion of the grain for a very long time, up to  tens of millions
   collisions,   it was
  found that no statistically significant result could be obtained
  with time-averaging method. The reason is fairly evident: with
  growing mass the grain motion slowed down, and it took more and more time
  for the grain to migrate from one   energy sub-band to another.

\begin{figure}
\includegraphics{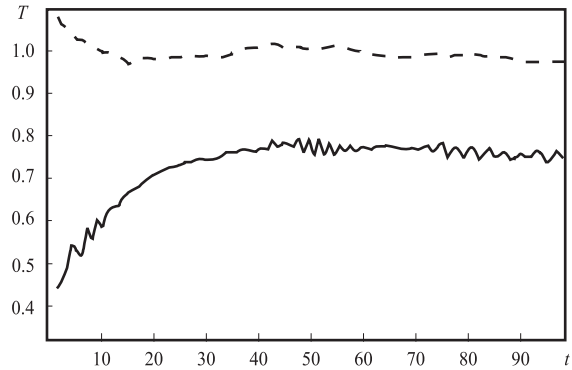}
\caption{\label{fig2}Temporal dependence of translational (solid
line) and rotational (dashed line) temperatures. }
\end{figure}

 The ensemble averaging yielded more meaningful results. Several typical energy
  distributions in logarithmic scale are depicted in Fig.~\ref{fig1}.
   The fit curves there
  correspond to Maxwellian distributions, {i.e.,} $F(E)\propto
  \sqrt{E}\exp(-E/T)$.
  Fig.~\ref{fig2}
 shows the temporal evolution  of translational (solid line) and
 rotational (dashed line) temperatures. The initial temperatures
 are  $T_{t0}=0.8\, T_0$ and $T_{r0}=1.3\, T_0$. The time in
 Fig.~\ref{fig2}
 is measured in the units of characteristic time of the  mass growth, $\tau_M$.
  According to
 Eq.~(\ref{mu}), $\tau_M=M_0/(4 \pi a_0^2 J)$, where $M_0$ and
 $a_0$ are the initial  grain mass and radius, respectively. The
 averaging was performed over the ensemble consisting of 1000
 samples.

 There was the sharp drop in translational temperature evolution at the
beginning of the computation. It originated due to the parabolic
potential well, in which the grain was kept. Initially the grain
was situated at the  bottom of the well, so its potential energy
was zero. Then, according to the virial theorem, half of its
kinetic energy transferred to the potential one. Since the
potential well did not influence the grain rotation, there was no
rotational energy drop.

Eventually the dust translational temperature tends to $\approx
0.75\; T_0$, that is, near the value given by Eqs.~(\ref{trm}).
The rotational temperature is also close to the value
(\ref{rotm}). Since there was only one  small grain, the
distribution of projectile atoms was essentially Maxwellian and we
were unable to reproduce the solutions (\ref{tra}) and
(\ref{rota}).

The simulations with the ``cold grain'' also confirmed the
analytical solution: the translational temperature tends to $2
T_0$ while the rotational temperature is always growing.

\section{Conclusion}

Although the simple model of the atom-grain inelastic collisions
accepted in this paper ignores some essential processes, it
demonstrates many interesting features. We confirmed that the
translational temperature of the dust component differs from the
temperature of the ambient gas. However, we demonstrated that the
``equilibrium'' temperature is highly sensitive to the details of
the inelastic collision. In particular, taking into account the
grain mass growth results in appreciable reduction of the dust
temperature. It should be noted that the difference between
various collision laws discussed above numerically is very small.
The discrepancies in energy balance of the order of a fraction of
a tiny mass ratio are accumulated and eventually result in
considerable effect. It is also worth pointing out that the
rotational temperature is sensitive even to the details of the
inner structure of grain.

The important lesson, which may be drawn from the above
discussion, is that there is no thermodynamic equilibrium between
dust and ambient gas. The statement itself is fairly evident since
a dusty plasma is an open system. However, this indicates the
inapplicability of the fluctuation-dissipation  theorem, which is
the basement of the Langevin approach to the theory of Brownian
motion. Therefore, the problem of deducing the Langevin equation
applicable to dusty plasmas arises.

\begin{acknowledgments}
This work was performed under the financial support granted by the
Sfb 555 of the Deutsche Forschung Gemeinschaft and the Netherlands
Organization for Scientific Research (NWO), grant \# 047-008-013.
One of us (A.M.I.) also acknowledges the support from Integration
foundation, project \# A0029. \end{acknowledgments}


\begin{thebibliography}{xx}
\bibitem{quinn1}R. A. Quinn and J. Goree, Phys. Plasmas \textbf{ 7}, 3904
(2000).
\bibitem{quinn2}R. A. Quinn and J. Goree, Phys. Rev. E \textbf{ 61} 3033 (2000).
\bibitem{zagor}
A. G. Zagorodny, P. P. J. M. Schram and S. A. Trigger, Phys. Rev.
Lett. \textbf{ 84}, 3594 (2000).
\bibitem{schram} P. P. J. M. Schram, A. G. Sitenko, S. A. Trigger and A. G.
Zagorodny, Phys. Rev. E \textbf{ 63}, 016403 (2001).
\bibitem{tsyt1}V. N. Tsytovich and  U. de Angelis,  Phys. Plasmas \textbf{ 6},1093 (1999).
\bibitem{tsyt2}V. N. Tsytovich and O. Havnes, Comments Plasma Phys. Control.
Fusion \textbf{ 15}, 267 (1993).
\bibitem{tsyt3}V. N. Tsytovich, U. de Angelis, R. Bingham and D. Resendes,
Phys. Plasmas \textbf{ 4}, 3882 (1997).
\bibitem{sitenko}A. G. Sitenko, A. G. Zagorodny, Yu. I. Chutov, P. P. J. M.
Schram, and V. N. Tsytovich, Plasma Phys. Controlled Fusion
\textbf{ 38}, A105 (1996).
\bibitem{ignatov} A.M.Ignatov,  Plasma Phys. Rep. \textbf{ 24}, 677 (1998).
\bibitem{childs} M. A. Childs and A. Gallagher, J. Appl. Phys. \textbf{ 87}, 1076 (2000).
\bibitem{gal} A. Gallagher,  Phys. Rev. E  \textbf{ 62}, 2690 (2000).
\bibitem{stoffels} W. W. Stoffels, E. Stoffels, G. H. P. M. Swinkels, M. Boufnichel
 and G. M. W. Kroesen,  Phys. Rev. E  \textbf{ 59}, 2302 (1999).
 \bibitem{13} A.M.Ignatov and S.A.Trigger, \eprint{http://arxiv.org/abs/physics/0006072}.
\bibitem{14} S.A.Trigger, Contrib. Pl .Phys \textbf{ 41}, 331 (2001) .
\bibitem{15} A.M.Ignatov, S.A.Trigger and W.Ebeling, Phys.Lett. A,
submitted, (    2001).
 \bibitem{jellum} G. M. Jellum, J. E. Daugherty, and D. B. Graves, J. Appl.
Phys. \textbf{ 69}, 6923 (1991).
\bibitem{am} A. M. Ignatov and Sh. G. Amiranashvili, Phys. Rev. E  \textbf{ 63}, 017402
(2001).
\bibitem{morfill} G. E. Morfill, H. M. Thomas, U. Konopka,
 H. Rothermel, M. Zuzic, A. Ivlev, and J. Goree, Phys. Rev. Lett.
\textbf{ 83}, 1598 (1999).
\bibitem{khrapak} 3.  B. M. Annaratone, A. G. Khrapak, A. V. Ivlev,
 G. Sollner, P. Bryant, R. Sutterlin,
 U. Konopka, K. Yoshino, M. Zuzic, H. M. Thomas, and G. E. Morfill,
 Phys. Rev. E \textbf{  63}, 036406 (2001).
\bibitem{ebeling} U.Erdmann, W.Ebeling, and V.Anishchenko,
Phys. Rev. E,  submitted (2001) .
\bibitem{wu} G. Shanmugam and V. Selvam, Phys. Rev.  C \textbf{62},
014302 (2000).
\bibitem{klim} Yu. L. Klimontovich, \textit{ Statistical physics}
( Harwood Academic  Publishers, New York, 1986).
\end{thebibliography}
\end{document}